\documentclass[twocolumn,showpacs,eqsecnum,amsmath,amssymb,aps,prb,floatfix]{revtex4-1}

\usepackage{graphicx} 
\usepackage{dcolumn} 
\usepackage{bm} 
\usepackage{hyperref} 
\usepackage{slashed}
\usepackage{subfigure}

\begin{document}

\title{Holstein-Hubbard model at half filling : A static auxiliary field study}

\author{Saurabh Pradhan}
\email{pradhanphy@gmail.com}
\author{G. Venketeswara Pai}
\affiliation{Harish-Chandra Research Institute, Chhatnag Road, Jhunsi, Allahabad 211019, India}


\begin{abstract}
We study the Holstein-Hubbard model at half filling to explore the ordered
phases such as the charge density wave and antiferromagnet. The Coulomb
interaction is rewritten in terms of auxiliary fields. By treating the
auxiliary fields and phonons as classical, we obtain real space features of
the system and transition between the phases from weak to strong
coupling. When both interactions are weak, mutual competition between
them leads to a metallic phase in an otherwise insulator dominated phase
diagram. Spatial correlations induced by thermal fluctuations lead to
pseudogap features at intermediate range of coupling.

\end{abstract}

\maketitle


\section{Introduction}
Several materials, notably transition metal oxides \cite{imada}, have strong Coulomb interactions among their constituent  electrons \cite{emery}, as well as strong coupling between
electrons and  the  underlying lattice\cite{holstein}. Interplay of such competing many body interactions often leads to  the emergence of effective energy scales and various
broken symmetry phases and transitions among them,  giving rise to significant changes in their low-energy behavior\cite{lee,freericks11}.
Understanding the combined effect of these two is a challenging problem and there has been some remarkable progress in the last couple of
decades.  Some of the systems that are known to have  both these interactions playing a role include high-$T_c$ superconducting cuprates\cite{lanzara,Dagotto,damascelli,egami},  alkali doped fullerides\cite{takab,capone,gunnersen}, bismuthates\cite{taraphder,bismuth},
and most notably doped manganites\cite{Dagottobook,rama}.  In cuprates\cite{lanzara,mishchenko}, kinks observed in ARPES are believed to be features arising from strong electron-phonon coupling which also give
rise to prominent features in inelastic neutron scattering and tunneling.  The system also has strong electron-electron interactions as evidenced by the Mott insulating state of the parent compound. In fullerides, an antiferromagnetic phase stabilized by Coulomb interactions, evolves to an $s$-wave superconducting state and it is believed that phonon effects are likely to be present. In doped bismuthates, a charge density wave\cite{gruener} transforms to an $s$-wave superconducting state upon doping; valence skipping arising due to Coulomb interactions and coupling  of charge carriers to breathing mode phonons are believed to be responsible for the behavior. Manganites\cite{dagotto1,millis3,millis1} present the most compelling case where orbitally degenerate electrons experience strong Mott-Hubbard interactions and are also coupled to octahedral symmetry lifting Jahn-Teller phonon modes. It is being realized that the conventional way of
treating only one of the interactions is inadequate for a proper understanding of these materials.\\
~\\
The Holstein-Hubbard model \cite{nagaosa,freericks,werner,fehske, Berger, Hewson, Johnston, freericks2,khatami, macridin,marsiglio, nowadnick, scalettar,vekic}
is the simplest starting point to theoretically explore the combined effect of these two interactions.  It describes a single-band electron coupled to an Einstein phonon mode. The Coulomb interaction is modeled by an on-site Hubbard term capturing the energy cost when two electrons of opposite spins are 
present at a given site. In real systems, this model could be an oversimplification. There could be multiple orbitals relevant\cite{goodenough}  as in manganites, leading to inter- and 
intra-Coulomb matrix elements. There could also be multiple phonon modes involved as happens in Jahn-Teller systems\cite{kanamori}. However, general features, leaving out specifics such as orbital ordering,
would be very well captured by the simplest model itself. For example, on a two-dimensional square lattice at half filling, the Hubbard interaction is expected to give rise to
a weak-coupling spin density wave transforming to a local-moment antiferromagnetic Mott-Hubbard insulator (MHI)  at strong coupling accompanied by a metal-insulator transition
at finite temperatures\cite{borejsza1,borjesza2}. On the contrary, the Holstein interaction promotes coexisting charge density wave and superconducting ground states, if phonon dynamics is retained. However, for static phonons, it is expected that a weak coupling charge density wave  would crossover to a bipolaronic insulator at strong coupling. Obviously, these phases will compete strongly when both interactions are  present. Motivated by this, there have  been several studies in recent years. These include analysis of various aspects of the problem
using Migdal-Eliashberg theory\cite{Berger}, quantum Monte Carlo (QMC) \cite{Berger,huang}, exact diagonalisation\cite{dobry1,dobry2}, variational treatments\cite{grilli} such as the Gutzwiller approximation\cite{barone} for  correlation, and dynamic mean field theory (DMFT) \cite{werner,diciolo,bauer1,Hewson,wernermillis, capone1, capone3}.  In particular, Bauer and Hewson\cite{Hewson} studied the ground state of the model at half filling using DMFT\cite{georges,freericks2,capone2,kollar,kollar2} in conjunction with numerical re-normalization group (NRG)\cite{wilson,pruschke}. A recent study\cite{werner} using dynamical mean field theory with continuous-time quantum Monte Carlo as an impurity solver has brought out several interesting features . These include strong renormalization of superconducting $T_c$ and the emergence of a paramagnetic metallic phase in the weak-coupling limit. While DMFT is by far one of the most reliable tools to study strongly correlated systems, it has certain limitations. It is exact in infinite dimensions or when coordination number is large; however the theory being a local one does not capture the full real-space features. If the system has geometrical constraints or frustration, a local theory will not be able to shed light on features intrinsic to them. It is also not possible to include
disorder in any meaningful way since crucial  features of interference cannot be captured in a single-site theory. Some new techniques are needed to overcome these
problems and complement DMFT in the cases mentioned above. While such methods too  will have their own limitations and  range of applicability, they may be able to explore systems that DMFT cannot handle, especially when they are bench-marked  in known cases. We use such a method to explore the problem at hand.\\
~\\
The method\cite{hubbard,schultz,bartosch,borejsza1,borjesza2,zaleski,rama2,tiwari} includes rewriting the quartic fermion interaction in terms of auxiliary fields corresponding to charge and spin degrees of freedom. However, the resulting problem is still a many-body one, albeit with new fields. To simplify matters, we concentrate on the static part of these fields\cite{rama2,tiwari} and also assume that the phonons are static.
This results in a problem of a single-band electron moving in the background of three classical fields: the charge and magnetization auxiliary fields and lattice displacements. At any temperature, the  statistically significant configurations of classical fields can be sampled employing a Monte Carlo (MC) procedure\cite{Dagotto,kumar}. The electron problem can be solved by exact diagonalization. The method captures both weak and strong coupling regimes as described in Sec. III.\\
~\\
We find that when only the Hubbard interaction is present, the system evolves from a Slater\cite{slater} to a Mott insulator (MI) \cite{imada} with nonmonotonic variation of the N\'{e}el temperature.
When only the lattice coupling is present, it transforms from a weak-coupling charge density wave\cite{gruener} to a bipolaronic insulator at strong coupling. When both are present, a critical line separates the two phases. At finite temperatures, the disordered phase appears to be metallic at weak coupling, but insulating at strong coupling. However, at intermediate coupling significant pseudogap features appear\cite{dupuis1} in the spectral function that modifies response of the electronic system such as optical transport in a significant way\\
~\\
The paper is organized as follows. In Sec. II, we describe the model in detail and the method employed. Section III is devoted to benchmarking with previous studies when only one of the interactions is present. In the next section, we give the ground-state (low temperature) phase diagram of the model resulting from the present study, followed by a detailed finite-temperature analysis of the electronic properties. Finally we conclude, describing the limitations of the method, advantages it has, and spell out future plans. 
 
 \section{Model and the static auxiliary field method}
As mentioned earlier, we look at the simplest model of a one-band electronic model coupled to a single-mode Einstein phonon with the Coulomb interaction assumed to be local on a
 two-dimensional square lattice.
The Hamiltonian is given by
\begin{eqnarray}
H &=& H_{tb} + H_{Hubbard} + H_{el-ph} + H_{ph}, \nonumber\\
H_{tb} &=& -t \sum_{\left< ij \right>, \sigma} c^{\dag}_{i\sigma} c_{j \sigma} + H.c., \nonumber\\
H_{Hubbard} &=& U\sum_i n_{i \uparrow} n_{i \downarrow}, \nonumber\\
H_{ph} &=& \sum_i {{{\bf p}^2} \over {2m}} + {K \over 2} \sum_i {\bf Q}^2, \nonumber\\
H_{el-ph} &=& g \sum_i  n_i   {\bf Q}_i.
\end{eqnarray} 

Here $H_{tb}$ is the kinetic energy of the electronic system with $t$ being the hopping parameter, $c_i$ being an electron destruction operator at site $i$, and $\left< ij \right>$ 
representing the nearest neighbors $j$  of site $i$. $U$ is the on-site Hubbard interaction and $g$ is the Holstein electron-phonon coupling. $H_{ph}$ is the Hamiltonian
for the Einstein phonon with frequency $\omega = \sqrt{K/m}$. Since we are interested in half filling $\left< n_i \right>=$ 1 .
Since the classical single-site Holstein Hamiltonian has a polaronic minimum with a distortion $\rho = (g/K)$, and polaronic binding energy $E_{pol} = - \left(g^2 /2K \right)$,
we scale the phonon coordinate $Q$ by $\rho$ and phonon energies by $\left| E_{pol} \right|$. This results in a single dimensionless parameter (scaled in terms of energy unit $t$)
for the phonon part of the Hamiltonian which we denote as $V$. From now on, we denote the dimensionless Hubbard interaction (in units of $t$) as $U$.
We shall explore the physics of this model as functions of these two dimensionless parameters.\\
~\\
To simplify this many-body problem, we perform  a Hubbard-Stratanovich (HS)\cite{schultz,hubbard} transformation of the quartic interaction term by introducing
two auxiliary fields, one each for the charge and magnetization sectors,. The scalar-valued charge auxiliary field at each site is $\phi_i(\tau)$ and the vector-valued magnetization auxiliary field is ${\bf m}_i (\tau)$.  
Since $ n_{i \uparrow} n_{i \downarrow} = n^2_i /4- {\left( {\bf s}_i \cdot {\bf m}_i \right)}^2$, where  ${\bf s}_i = {1 \over 2} \sum_{\alpha,  \beta} c^\dag_{i \alpha} \vec{\sigma}_{\alpha \beta} c_{i \beta} $,  we can write

 \begin{equation} 
 e^{U n_{i \uparrow} n_{i \downarrow}}  = \int {{d \phi_i d {\bf m}_i} \over {4 \pi^2 U}} \exp \left( {\phi^2_i \over U} + i \phi_i n_i + 
 {{\bf m}^2_i \over U} - 2 {\bf m}_i \cdot {\bf s}_i \right).
 \end{equation}

This results in a quadratic fermion problem in which fermions move around
in a (quantum-mechanical, time-dependent) background of the two auxiliary fields and the phonon field which is computationally, still, a challenging problem.
The partitions function is given by

 \begin{eqnarray}
  {\cal Z} &=& \int   \Pi_i  { {dc^\dag_i dc_i d \phi_i d {\bf m}_i} \over {4 \pi^2 U}}  dQ_i  \exp \left( - \int_0^\beta  d \tau {\cal L} \left( \tau \right) \right), \nonumber\\
  {\cal L} \left( \tau \right) &=&  \sum_{i,  \sigma} c^\dag_{i \sigma}  \left( \tau \right) \partial_\tau c_{i \sigma} \left( \tau \right) 
  - t \sum_{\left< ij \right>, \sigma} c^\dag_{i \sigma} c_{j \sigma}  \nonumber\\
  &+&  {\cal L}_{cl} \left( \phi_i \left( \tau \right), {\bf m}_i \left( \tau \right) \right) + {\cal L}_{ph},
  \nonumber \\
  {\cal L}_{cl}   &=&  \sum_i \left[ {\phi^2_i \over U} + i \phi_i n_i + {{\bf m}^2_i \over U}
  - 2 {\bf m}_i \cdot {\bf s}_i \right],
  \end{eqnarray}
where ${\cal L}_{ph}$ is the phonon Lagrangian.\\
~\\

We make the following approximations. We assume that all three background fields are classical and hence neglect their time dependence. We retain their spatial dependence
and do a thermal averaging  of their configurations at every temperature numerically. We limit ourselves to half filling, i.e., one electron per
site, in this paper. In this spirit, we make a saddle-point approximation for the static charge field, i.e., $\phi_i \rightarrow \langle\phi\rangle = (U/2) \langle n_i\rangle  = U/2 $, and this is taken to be
site independent.  Upon rescaling  ${\bf m}_i \rightarrow \left(U/2 \right) {\bf m}_i$, the resulting Hamiltonian reads :\cite{borejsza1,tiwari}
\begin{eqnarray}
H_{eff} &=& -t \sum_{\langle ij\rangle, \sigma} c^\dag_{i \sigma} c_{j \sigma} - \mu_{eff}N  -{U \over 2} \sum_i {\bf m}_i \cdot \vec{\sigma}_i  \nonumber\\
&+&  {U \over 4} \sum_i {\bf m}^2_i  
+ V\sum_i n_i   Q_i + V \sum_i  Q^2_i,
\end{eqnarray}
 where $\mu_{eff} = \mu - U/2$ with  the partition function being given by

 \begin{equation}
{\cal Z} = \int {\cal D} {\bf m}  {\cal D}Q   {\cal D} {\left[ c^\dag , c  \right]}  \exp \left( - \beta H_{eff} \right).
\end{equation}

For a given configuration of $Q_i$ and ${\bf m}_i$, the Hamiltonian (quadratic in fermions) needs to be diagonalized just once. However, one needs to sample most probable configurations of both $Q_i$ and ${\bf m}_i$ at every temperature and they have to be determined from corresponding distributions :
\begin{eqnarray}
P \left( Q_i  \right) &=&{ { \int {\cal {D}} {\bf m} { {\cal D} \left[ c^\dag, c \right]  e^{-\beta H_{eff}}}}  \over {\int {\cal {D}}{\bf m} {\cal {D}}Q 
  { {\cal D} \left[ c^\dag, c \right]  e^{-\beta H_{eff}}} }}, \\
  P \left( {\bf m}_i \right) &=&{ { \int {\cal {D}} {Q} { {\cal D} \left[ c^\dag, c \right]  e^{-\beta H_{eff}}}}  \over {\int {\cal {D}}{\bf m} {\cal {D}}Q 
  { {\cal D} \left[ c^\dag,  c \right]  e^{-\beta H_{eff}}} }}.
  \end{eqnarray}

 While it appears that  the neglect of the time-dependent effects reduces this method to unrestricted Hartree-Fock for the ground state, it retains the full  classical thermal fluctuations in an unbiased way which leads to significant changes from HF results at finite temperature and smoothly interpolates between known limits at weak and strong coupling.\\
  ~\\
 The probability distribution functions appearing above are not exactly calculable since they involve tracing over fermions and integrating over all static configurations of the classical fields. We generate the equilibrium configurations for the classical field self-consistently using a Monte Carlo method\cite{dagotto1}. This is achieved by starting with a given set of configurations, and     
  attempting an update which requires diagonalizing the fermion Hamiltonian and generating  most probable configurations using the standard MC method. However, this severely
  restricts the system size of the problem, even though the fermionic part is quadratic. To explore higher system sizes, we use a traveling cluster algorithm\cite{kumar}, in which a small cluster 
  around the reference site is diagonlized and energy cost evaluated for MC update. During the MC procedure, as the reference site keeps moving on the lattice, so does the cluster.
  The results presented in this paper employ a cluster size of 8  $\times$ 8 and the largest system size used is 32 $\times$ 32.
  Once the system reaches equilibrium, we evaluate thermal averages of structure factor for charge density and magnetization. 
  \begin{eqnarray}
  N\left( {\bf q} \right)  &=& {1 \over N^2} \sum_{ij} \left< n_i n_j \right> e^{i {\bf q} \cdot \left( {\bf r}_i - {\bf r}_j \right)}, \\
  S\left( {\bf q} \right)  &=& {1 \over N^2} \sum_{ij} \left< {\bf m}_i   \cdot  {\bf m}_j \right> e^{i {\bf q} \cdot \left( {\bf r}_i - {\bf r}_j \right)}. 
   \end{eqnarray}
  Spectral and transport properties for the fermion system have also been evaluated in thermal equilibrium which is described in Sec. V.
  
 \section{Exploring the Hubbard and Holstein Physics}
 In this section, we  present the phase diagram of the model for the individual cases when either the Holstein term is absent (the Hubbard model) or the Hubbard
 term is absent (the Holstein model). Both these problems have been studied extensively in the past and it will help us benchmark our results.\\
 ~\\
 When the Holstein term is absent, the model reduces to a single-band Hubbard model on a two-dimensional square lattice at half filling. This does not have a metallic ground state\cite{bartosch,borjesza2,dupuis1} for any nonzero value of $U$ and has long-range antiferromagnetic insulating (AFI) order in the ground state. For small $U$, a Slater instability results due to nesting of the Fermi surface  and the system is a spin density wave with a gap in the spectrum. For large $U$, the physics of superexchange takes over, due to the "no double occupancy constraint" and the resulting kinetic energy reduction due to virtual hopping. The system is a Mott-Hubbard insulator with local moments present 
 whose low-energy properties are governed by the antiferromagnetic Heisenberg model.  The magnetic transitions  resulting from these two behaviors have very different $U$ dependence. 
 At small $U$ the  $T_N$ scales with $U$ as expected in an unrestricted HF treatment and results in a paramagnetic metallic phase (PM) above $T_N$ due to the closing of the Slater gap . However,  at large $U$,  $T_N \sim (1/U)$ due to Mott physics and results in a paramagnetic insulator (PI). The present method captures both these behaviors very well. The finite-temperature phase diagram also looks qualitatively different from the HF phenomenology. While for small $U$, the Slater gap closes at $T_N$, there is a pseudogap (PG)  state that appears at intermediate values which crosses over to a paramagnetic Mott-Hubbard  insulating state at large 
 $U$. The paramagnetic state has strong AF fluctuations, especially in the intermediate range of the coupling constant, which results in pseudogap features in the spectral function.\\
 ~\\
 We now consider the case when the Hubbard term is absent, resulting in the half-filled Holstein model on a two-dimensional square lattice. This model again does not have a metallic ground state for any nonzero $V$. For small $V$, there is a Peierls instability leading to a charge density wave due to nesting, with a gap in the spectrum, which we call a charge-ordered insulator (COI). The charge modulation occurs at a wave vector $(\pi, \pi)$. At finite temperatures, the gap shrinks and vanishes at $T_{CDW}$ above which the system is a nonmagnetic metal (NMM) with passive spin degrees of freedom. As $V$ increases, the ground state of the system evolves through this charge-ordered state resulting in a bipolaronic insulator (BPI)  at very large $V$. This can be understood due to a mechanism similar to superexchange for the spins.  Since $U$ is absent, there is no energy cost for double occupancy and a bipolaronic state lowers the energy through virtual fluctuations of charge. In this limit the physics can be described using a nearest-neighbor-interaction model,  i.e., $H = \alpha \sum_{\langle ij \rangle} n_i n_j$, where $\alpha \sim (1/V)$ and hence the charge-ordering temperature goes as $(1/V)$. At intermediate values of $V$, a pseudogap phase intervenes which has spectral and transport features similar to the one previously mentioned. The finite-temperature, large-$V$ phase is insulating with charges remaining as bipolarons, but losing their long-range order.The spin degrees of freedom are passive in the entire phase diagram and the magnetization vanishes. We give the two phase diagrams in the $U-T$ and $V-T$ planes in Fig. 1.\\
 ~\\
 In passing, we wish to point out that the above results are indeed not exactly what is expected in two dimensions since there cannot be any finite-temperature transitions. These results should be taken as suggestive of what would happen in higher dimensions or as crossover scales where correlation lengths increase rapidly. (See the concluding section.) Further, we characterize the pseudogap phase  as one in which the density of states does not have any perceptible hard gap, but has a dip at the chemical potential, suggesting a dramatic decrease of the low-energy spectral weight.
 There is no real phase transition occurring here. It should be thought of as a crossover to a region where the density of state appears quite different from that of an insulator with a hard gap.\\
 ~\\

  \begin{figure}
\includegraphics[scale=0.5]{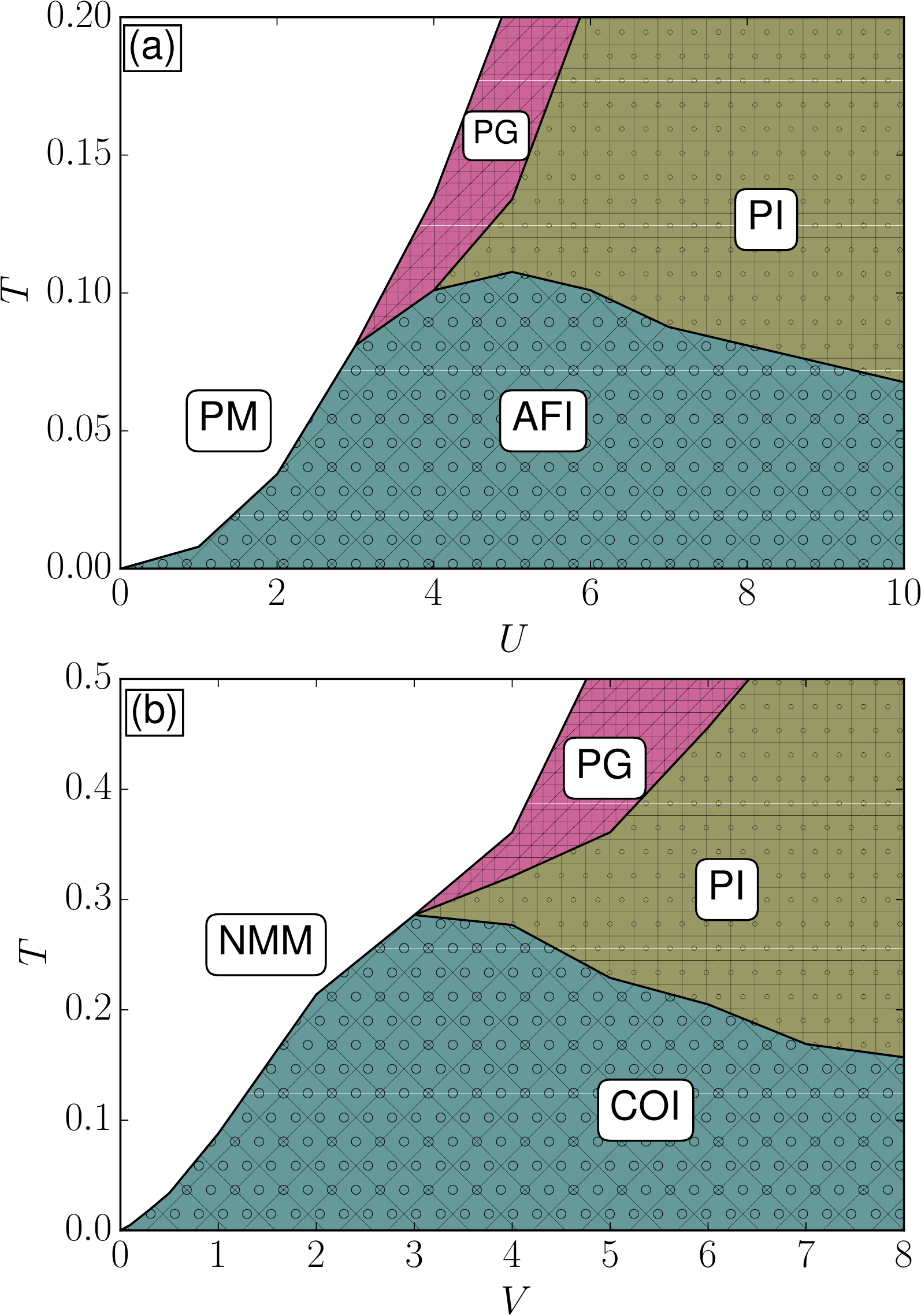}
\caption{  (Color online)  (a) The phase diagram of the Hubbard model on a two-dimensional square lattice at half filling. AFI, PM, PI, PG represent antiferromagnetic insulator, paramagnetic metal, paramagnetic insulator and pseudogap phases.   (b) The phase diagram of the Holstein model on a two-dimensional square lattice at half filling. COI, NMM represent charge-ordered insulator and nonmagnetic metal. }
\label{fig:fig1}\end{figure}
 
 \section{Ground state properties  and Phase Transitions of the Holstein-Hubbard Model}
 Having clarified the trends that one obtains for the Hubbard and Holstein interactions separately, we now proceed to discuss the results for the full problem.
 However, in this section, we will concentrate on the ground-state properties and the nature of the phase transitions at finite temperatures. This includes the $U-V$  phase diagram at $T=$ 0, spectral functions of the fermions, probability density functions for the lattice variables, and charge and magnetization field configurations in real space. The above information would help us correlate various trends and elucidate the physics that emerges. As the phases change while changing parameters, we will see that correlated changes occur in properties of the fermionic, phononic, and auxiliary field variables.\\
 ~\\
 In Fig. (2) we present  the ground-state phase diagram
 of the Holstein-Hubbard model as a function of $U$ and $V$. As expected the results along the horizontal and vertical axes (corresponding to cases when one of the parameters is absent) confirm the discussion in the previous section. The phase diagram is almost entirely dominated by insulating regions. This is not surprising since individually, each interaction tries to localize electrons giving rise to a band/Mott/bipolaronic insulator. In the intermediate\cite{Hewson,werner} to large values of the scaled parameters, there is a transition between  a charge-disordered magnetic insulator to a charge-ordered nonmagnetic insulator. For example, at large values of $U$, an otherwise MI in the absence of Holstein interaction transforms to a BPI as  $V$ increases. This is a result of the two competing interactions. While a large $U$ tries to localize individual electrons at every lattice site at half filling, the Holstein interaction develops bipolaronic instability as discussed in the last section. When the energy scales become comparable, the system develops an instability and moves from one to the other..Notice that the spin structure factor at $(\pi, \pi)$ is nonzero in the AF phase and  $S({\bf q}) =$ 0 in the BPI phase signaling a
 nonmagnetic state. Similarly, the charge structure factor has a peak at ${\bf q} =$ (0,0) in the AF phase, and the modulation vector changes to $(\pi, \pi)$ in BPI phase.  
 At intermediate values of 
 $U$ and $V$ this behavior  persists for both the structure factors but is much less pronounced compared to the strong-coupling limit. This is the crossover regime between the Slater-MHI due to Hubbard correlations and Peierls-BPI crossover due to Holstein interaction. Figure 3 depicts trends of both the structure factors as a function of $U$ and $V$ and confirms our conclusion about the transitions. Previous studies \cite{horsch} on the $t-J$-Holstein model have shown that  as the exchange $J$ decreases,  the critical electron-phonon coupling
 required for the transition from AFI to COI increases.  This is consistent with our results since  $J \sim 1/U$.
 However, there exists a thin sliver of window in the $U-V$ plane at low interaction strengths where the system is metallic. This is in contrast to the case where the system is insulating when only one of the interactions is present.  This behavior is exemplified in Fig. (3) where the structure factor at these values is plotted.  This metallic behavior has been observed in previous studies of this model employing  DMFT\cite{Hewson,werner} using continuous-time QMC and  NRG as impurity solvers.  This unexpected metallic phase results from the fact that while the nesting\cite{gruener,slater} at half filling in the two-dimensional tight-binding model supports magnetic or charge-ordering instabilities separately, the competing interactions have a destructive effect on the transition since it frustrates different degrees of freedom, {\it viz.}, charge and spin in our case.
 The energy gained by a small mean field gap opening up in either channel is not sufficient to lower the absolute ground state energy when the other channel is included. This phase, in fact, brings out the true competition between the two interactions, where one acts predominantly over the spin sector while the other over the charge sector.
 A previous DMFT study \cite{wernermillis} has revealed that this metallic region  expands to larger $U$ and $V$ values as phonon frequency increases,
which could explain the stability of this phase in the classical phonon limit. 
 Further, notice that the phase boundaries merge to zero values of both parameters in our case in contrast to DMFT results. This is easily understood since  our method preserves the nesting instability of the two-dimensional non-interacting electron system whereas methods such as DMFT ignore them.\\
 ~\\
  \begin{figure}
\includegraphics[scale=0.50]{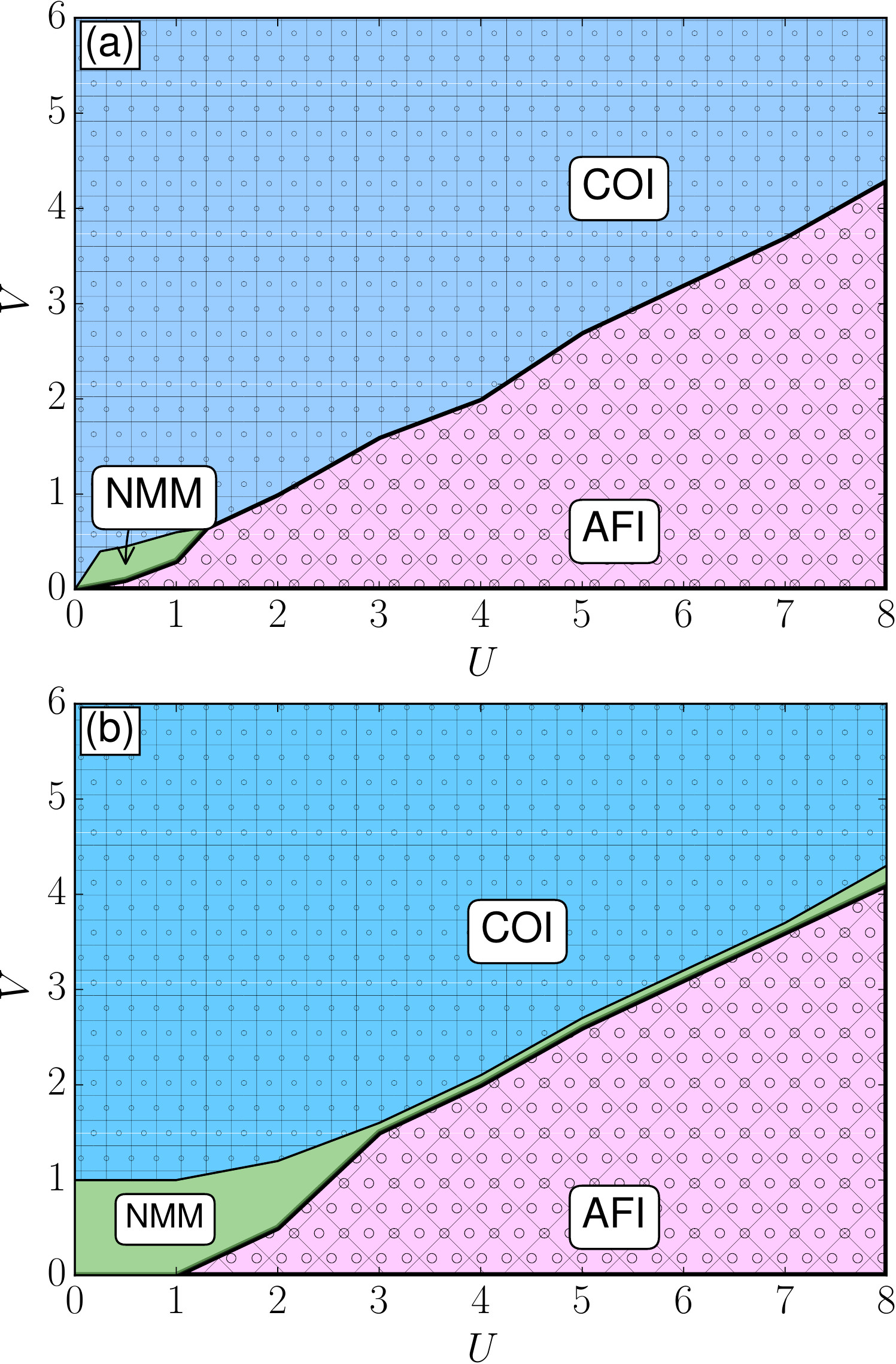}
\caption{(Color online) The ground-state and finite-temperature phase  diagram of the Holstein-Hubbard model as a function of scaled parameters $U$ and $V$. 
AFI, COI, NMM represent  antiferromagnetic insulating, charge-ordered insulating and nonmagnetic metal phases. The transition between AFI and COI is a weak first-order one. The temperatures are (a) $T =$ 0.001 and (b) $T =$ 0.050.}
\label{fig:fig2}
\end{figure}
 ~\\
 
  \begin{figure}
\includegraphics[height=5.50cm]{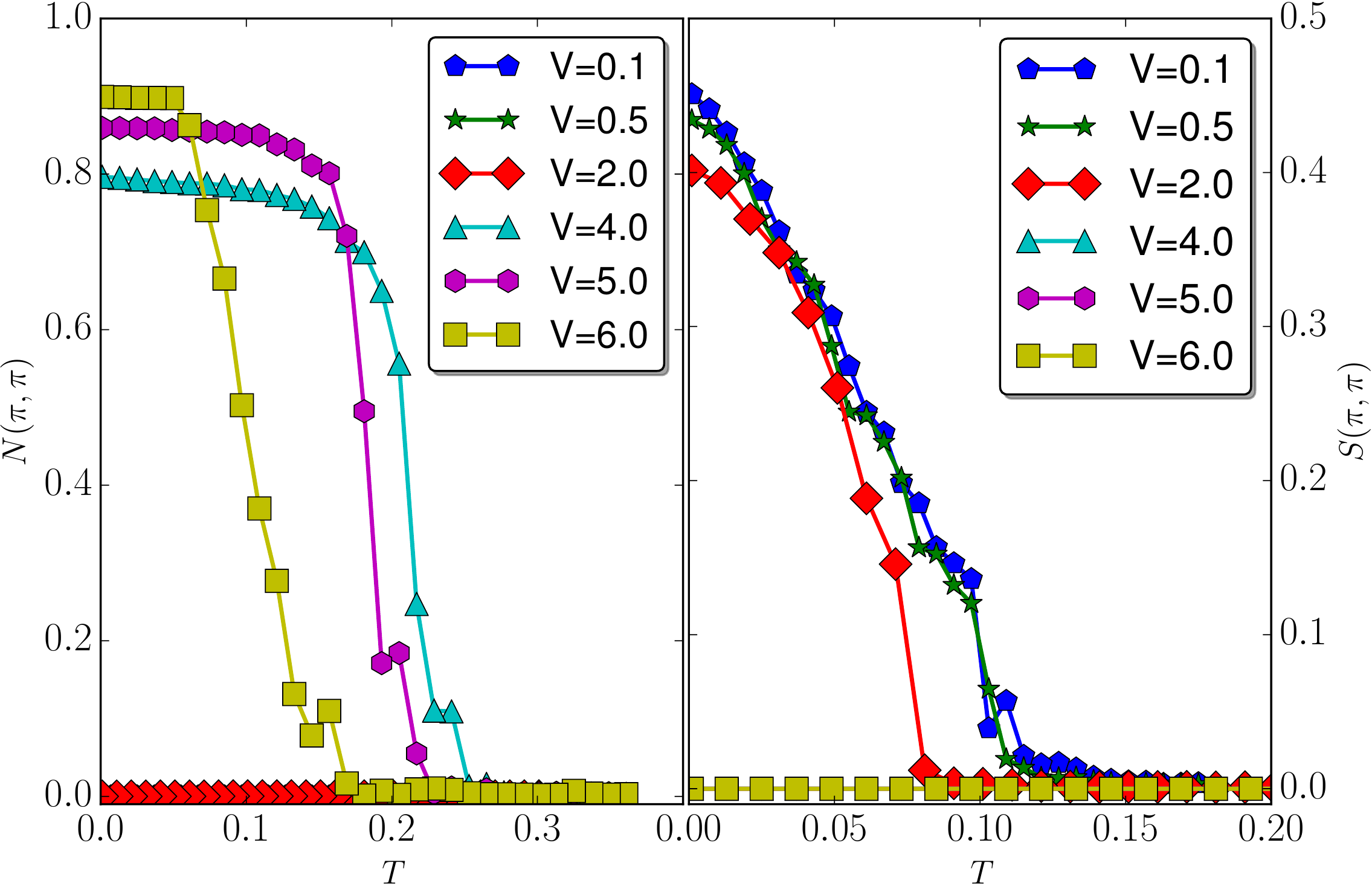}\\
\caption{(Color online) Structure factor corresponding to charge density wave $N(\pi,\pi)$(left) and the antiferromagnetic structure factor  $S(\pi,\pi)$(right) versus temperature for $U =$ 4.0 and some representative values of $V$. }
\label{fig:fig3}
 \end{figure}
 
 The MC procedure allows us to track the PDF of the phonon displacement variables across the transitions/crossover which is plotted in Fig. (4).  In the AF phases we see that 
 $P(Q)$ is a unimodal function peaked at $Q =$ -1, which implies that while every lattice site is distorted, it accommodates at most one electron per site. The distribution grows sharper as we grow from Slater to Mott limit, but the unimodal nature does not change. In this limit, Hubbard correlations play a larger role and the system tries to reduce the maximum number of electrons to one per site. At intermediate and strong coupling, at fixed $U$ as we increase $V$, we find that this unimodal distribution slowly crosses over to a bimodal one. This occurs because of the weakening of the Hubbard correlation and increasing role of the polaronic distortion energies. Two electrons of opposite spin occupying the same site lower the electron phone energy more and the system develops a bipolaronic instability [also see Figs. (2a) and (2b)]. In the nontrivial metallic phase, while every site is still distorted, the amplitude is very small.  These results, indeed, correlate with the charge structure factor and phonon probability distribution function.\\ 
  
   \begin{figure}
\includegraphics[height=7.0cm]{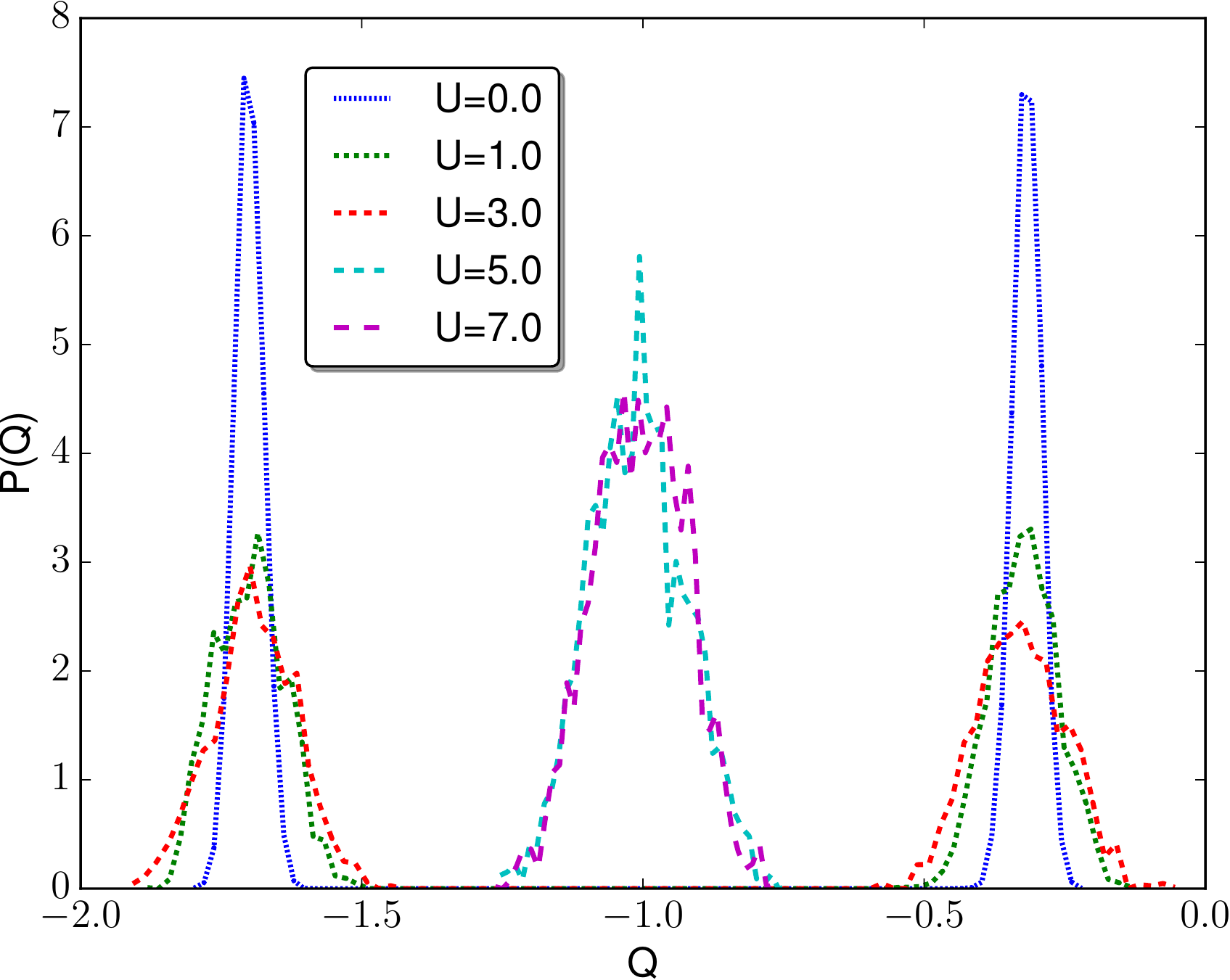}\\
\caption{(Color online) Phonon probability distribution for different values of $U$ at $V =$ 2.0, $T =$ 0.001. }
\label{fig:fig4}
 \end{figure}
 
 Our method allows us to provide a direct picture of the real-space correlations between the static magnetic auxiliary field, charge density, and  phonon variables at various sites.
 This will elucidate the character of the transition and especially the metallic phase that arises. In Fig. (5) we present snapshots of spin and charge over the lattice for a given set of parameters at a given instant of MC simulation after the system has equilibrated. Nonlocal correlations  among charge and magnetization fields can, in principle, be extracted from here. As expected, for lower values of $V$, spin correlations develop as $U$ increases moving to a local moment value in the MI phase. Such spin correlations are absent in the COI phase. On the contrary, charge densities modulate as $V$ is changed for a given $U$  resulting in a bipolaronic state.  In the corresponding phases spin modulation is negligible.  In the metallic phase, both densities remain negligible on average, but there are fluctuations.   The snapshots show a given configuration with some variation in the densities.
 However, an average over such configurations results  in uniform charge density and negligible magnetization confirming that it is indeed a metallic phase. \\
 ~\\
 
   \begin{figure}
\includegraphics[height=6.50cm]{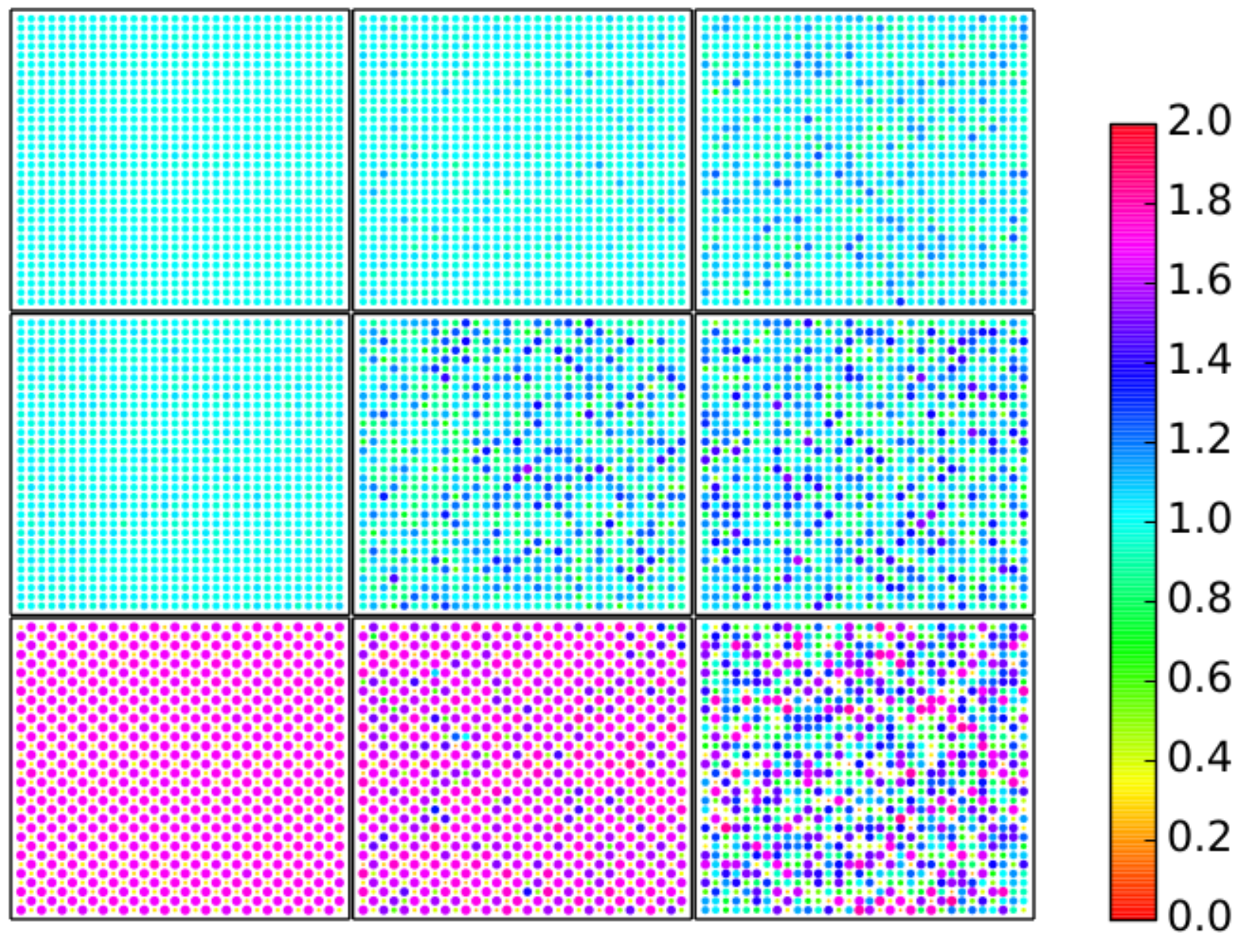}
\includegraphics[height=6.50cm]{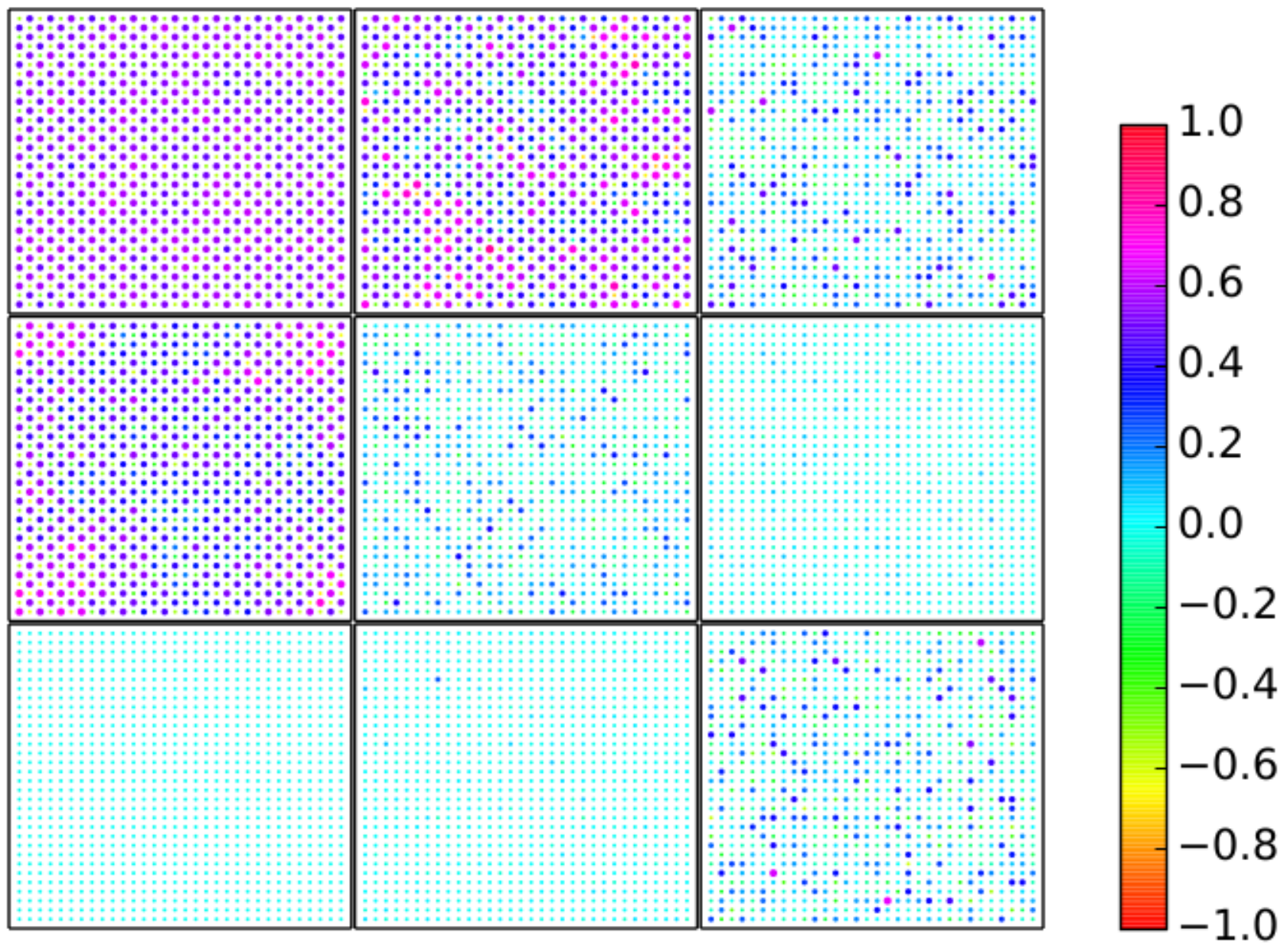}
\caption{(Color online ) Charge $n_i$ (upper)  and spin configuration $\mathbf{S_i.S_0}$ (lower)  for $U =$ 2.0. Temperature increases from left to right. Center column shows the configuration near the $T_c$. Three rows for $V =$ 0.50 (top), 1.0 (middle), 2.0 (bottom) and the system size is 32  $\times$ 32.}
 \label{fig:fig5}
 \end{figure} 
 
 The various physical quantities that we have used to characterize the ground-state properties confirm the expected behavior and is reflected in such diverse variables as charge, magnetization, distribution of lattice displacements,  and fermion spectral functions. The real-space picture gives a handle on how to correlate them. As will be discussed in the final section, this gives the added advantage of visualizing such changes in non-trivial geometries and especially on frustrated lattices, which is  intractable  or computationally expensive using other methods such as DMFT or its cluster variants. \\
 ~\\
 
 To conclude this section, within the static auxiliary field approximation of the decoupled HS fields that we have resorted to, we find a phase diagram that at low values of electron-phonon coupling crosses over from a Slater to MH insulator as $U$ is increased, and a Peierls to BPI as $V$ is increased for low values of Hubbard interaction. At intermediate to large values of coupling, there is a transition from antiferromagnetic MHI to a nonmagnetic  charge-ordered or bipolaronic insulator. However, there is a sliver of metallic phase at low coupling that results from frustrating effects of two interactions in different (charge and spin) channels. The behavior of different degrees of freedom correlates with these changes providing us with
 an efficient way to extract physics from weak to strong coupling.

 \section{Spectral and Transport Properties}
 Significant changes are expected in the phase diagram at finite temperatures due to the inclusion of "full"  thermal  fluctuations of the static field through configuration sampling.
 This was already noted in Sec. III where the effect of each interaction was looked at separately. In this section we present the results for various physical properties at finite temperatures and converge on the finite temperature phase diagram.\\
 ~\\
 Figure  6 shows the thermal-averaged single-electron spectral function $A(\omega)$ for different parameters at different temperatures. Deep in the insulating phase and at low temperatures they show a very clear gap and there are no states available at the Fermi energy as expected.  In the region where metallic ground state appears, on the contrary,  there is  nonzero spectral weight 
 at the Fermi energy even at the lowest temperatures. As temperature increases, we notice three regimes signifying different spectral features. For large values of $U$ and/or $V$, we find that the gap persists even for large temperature. This is due to the Mott-Hubbard or bipolaronic nature of the phases.  The fact that this feature survives at these values of parameters shows that the present method is capable of capturing the strong-coupling physics of this problem in both channels. At weak couplings, where a Slater or Peierls insulating phase is expected or the metallic phase emerges, the spectral features are very different at high temperatures. The gap vanishes entirely in the former cases and there is sufficient weight at the Fermi energy in all the three regimes. This clearly shows that the gap arises solely due to the nesting instability of the underlying Fermi system and the resulting order in either spin or charge channels. Once the order is destroyed, so is the gap. The most interesting features arise at intermediate values. Here a hard gap is not seen though there is significant reduction of spectral weight near the Fermi energy. There is spectral weight transfer from the coherence peaks to energies within the gap. This pseudogap feature arises due to persistence of local correlations in static fields even after the 
long-range order is destroyed. \\
 ~\\
 
 \begin{figure}
\includegraphics[height=6.0cm]{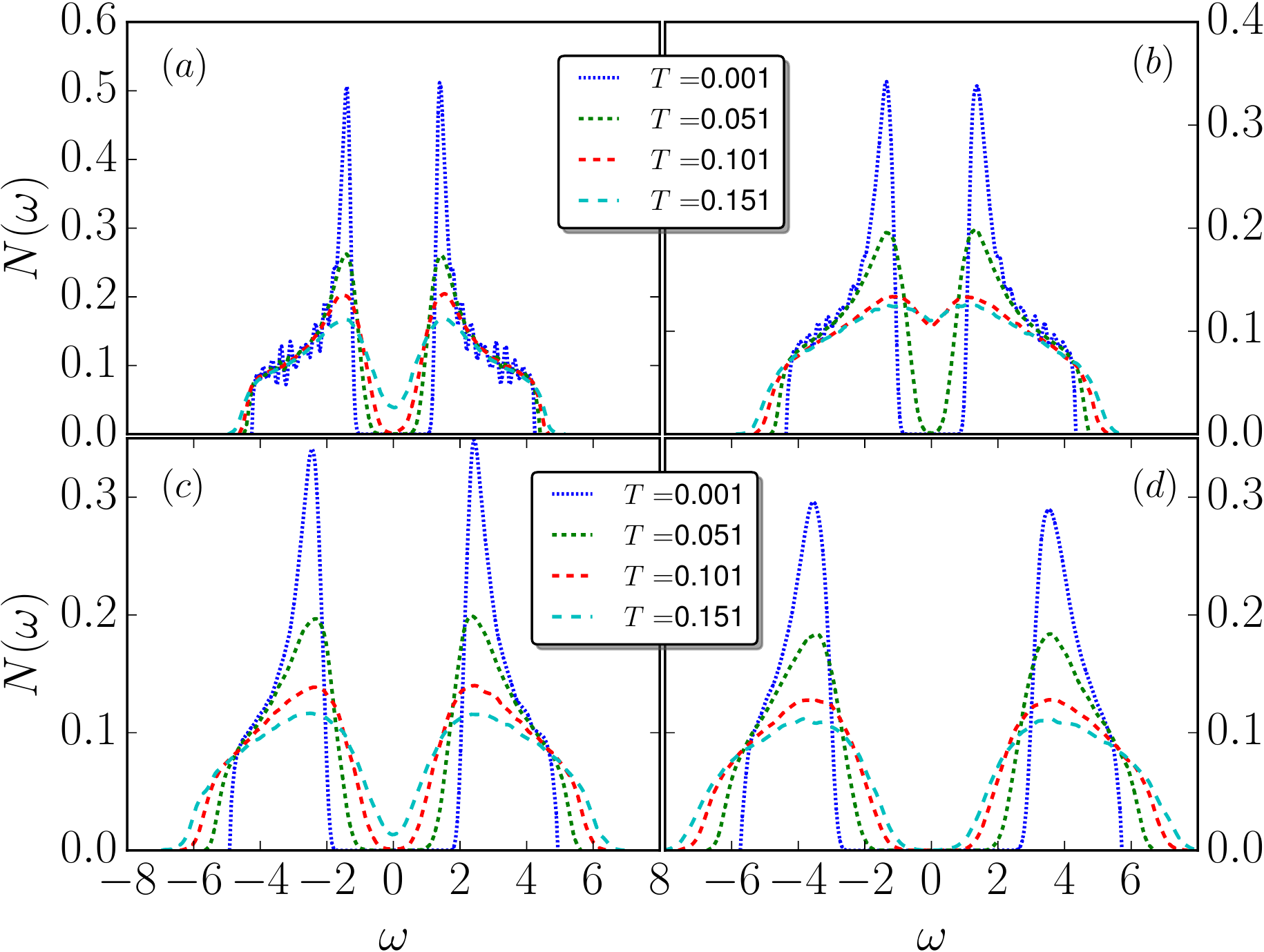}\\
\caption{(Color online) Density of state for different values temperature at constant  $V=$ 2.00 and $U$ varying across the charge density wave-antiferromagnetic transition. $U =$ 1.0 (a) , 4.0 (b), 6.0 (c), 8.0 (d).}
\label{fig:fig6}
 \end{figure} 
 
 The MC snapshots throw more light on the existence of short-range order in either spin or charge degrees of freedom at temperatures near or above the ordering temperatures. This is shown in Fig. (5). In each case, the states evolve from the ground states shown in Fig. (5). However, unlike the low-coupling counterparts, the local order persists even above transition temperatures. This local order, we believe, is the reason for the appearance of pseudogap-like features in spectral functions. However, unlike the strong-coupling cases, where a local moment or a bipolaron formation is favored and the spectrum shows a hard gap, the intermediate range does allow fluctuations in charge and spin variables at very site, leading to spectral weight appearing in the otherwise gapped region. We have verified that the phonon PDFs also exhibit persistence of bimodality in this region. \\
 ~\\
 
  \begin{figure}
\includegraphics[height=6.0cm]{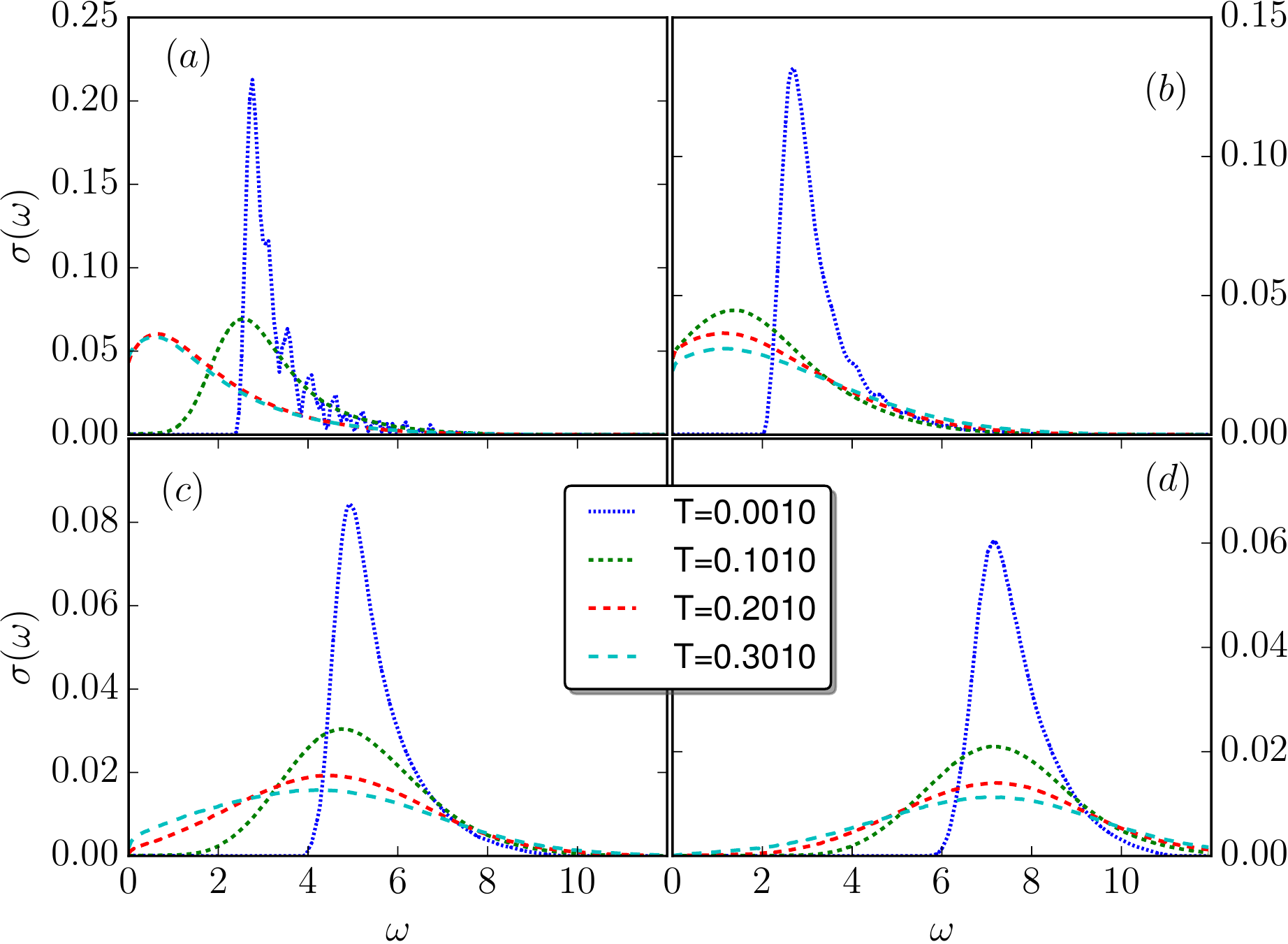}
\caption{(Color online) Temperature dependence of the  optical conductivity for $V =$ 2.0 and $U =$ 1.0 (a), 4.0 (b), 6.0 (c), 8.0 (d). }
\label{fig:fig7} 
 \end{figure} 
 
 The transport can be captured in an exact way without resorting to approximations as in cluster DMFT. To this end, we use Kubo formula\cite{tiwari} for the in-plane resistivity which involves the exact eigenvalues ($\epsilon_\alpha, \epsilon_\beta$)  and wave functions ($\left| \alpha \right>, \left| \beta \right>$)   of fermions obtained from diagonalization at several equilibrium configurations. 
 
 \begin{equation}
\sigma_{xx}  \left( \omega \right) = {\sigma_0 \over N} \sum_{\alpha, \beta} {{f\left(\epsilon_\alpha\right) - f\left( \epsilon_\beta \right)} \over {\epsilon_\beta - \epsilon_\alpha}}  {\left| \left< \alpha \left| J_x \right | \beta \right> 
\right|}^2
\delta \left( \omega - \left( \epsilon_\beta - \epsilon_\alpha \right) \right).
\end{equation}
Here $f$ denotes the corresponding Fermi function and the current operator $J_x$ is given by

\begin{equation}
J_x = -i  t\sum_{i, \sigma} \left( c^\dag_{i,\sigma} c_{i+x, \sigma} - h.c. \right),
\end{equation}
where $ \sigma_0 = {{\pi e^2} \over \hbar}$. The dc conductivity is obtained by letting $\omega \rightarrow  0$.\\
~\\

  Figure 7 shows the evolution of optical conductivity for a fixed value of $V$, but for varying $U$ at different temperatures. A notable feature is the non-Drude behavior of $\sigma\left( \omega \right)$.  Further, the pronounced low-frequency hump in the optical conductivity for small frequencies evolves into an inter-band Hubbard peak  as $U$ increases. A similar feature has been observed as we vary $V$ where the Hubbard peak gets replaced by the higher energy bipolaronic peak. The non-Drude behavior emanates from the pseudogap nature of the electronic spectral function that originates from strong local charge/spin fluctuations as discussed earlier. The $dc$ resistivity is plotted in Fig. (8) for a fixed value of $V$, but varying values of $U$ at different temperatures. A metal-to-insulator transition is clearly visible for weak-coupling regime ($U = 1$) in the inset.\\
 ~\\
 
\begin{figure}
\includegraphics[height=6.0cm]{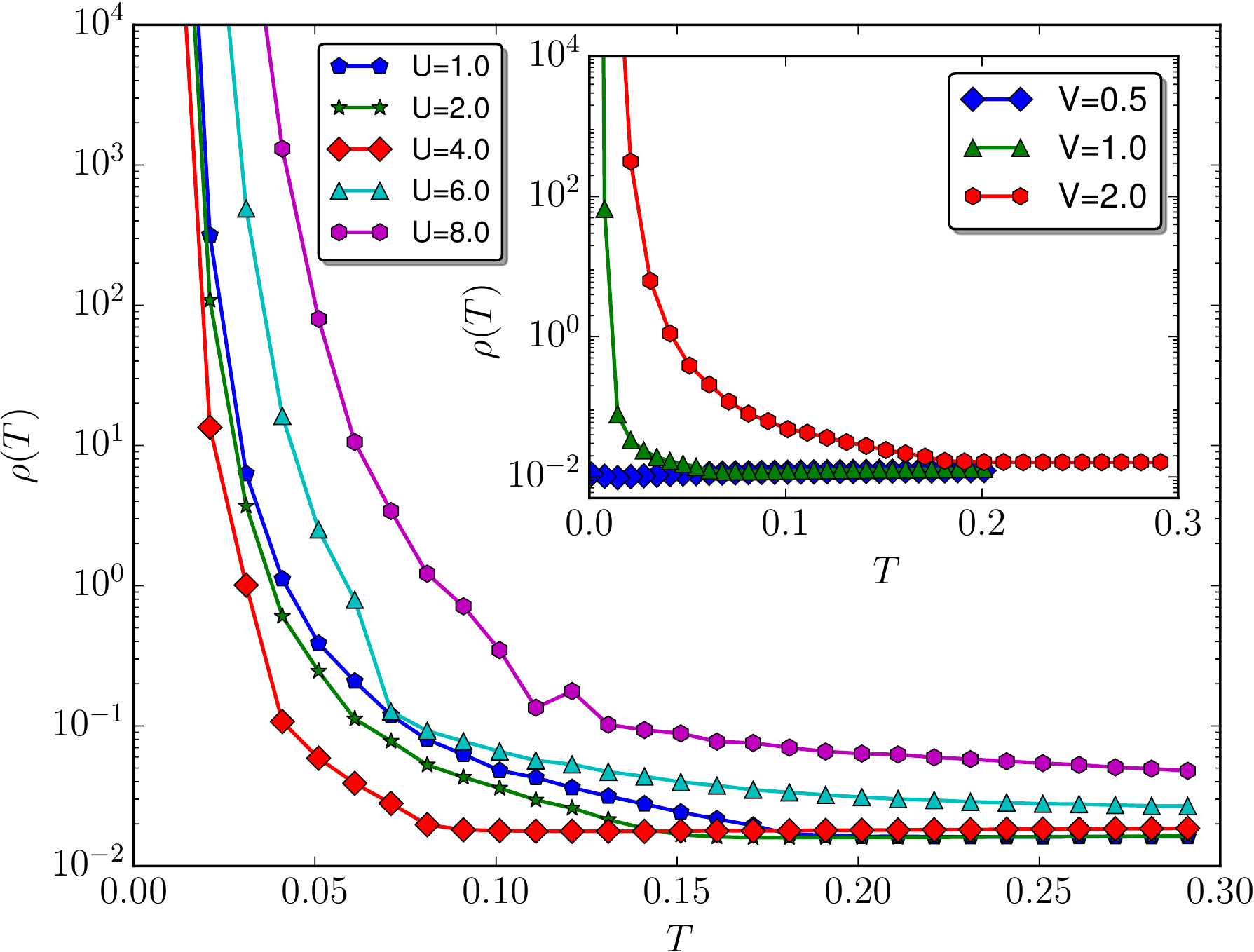}
\caption{(Color online) Temperature dependence of resistivity, $\rho(T)$, for various values of $U$ at $V =$ 2.0. Metal-insulator transition in the weak-coupling region for $U =$ 1.0 (inset).}
\end{figure}  
 
 Finally, we present the finite-temperature phase diagram of the model in Fig. (9). The phases include AF or CO insulating phases at low temperatures except for the sliver of metallic phase discussed earlier, metallic nonmagnetic phases at weak couplings, Mott-Hubbard and bipolaronic insulating phases at large couplings, and the pseudogap phase at intermediate coupling. The high-temperature behavior from metallic to insulating  is a crossover.  Note that we characterize the finite-temperature metallic phase by sign of  the temperature variation of the resistivity,  $d \rho /dT$.  It remains open as to how these instabilities would be affected due to quantum dynamics of the auxiliary field or phonons. However, the remarkable qualitative agreement with previous DMFT studies suggests  that the quantum dynamics of these fields may not be relevant for the regime we have concentrated on. Further, it appears that the current method may be used for geometries and systems where DMFT treatment may not be applicable as we discuss below.
 
  \begin{figure}
\includegraphics[height=6.0cm]{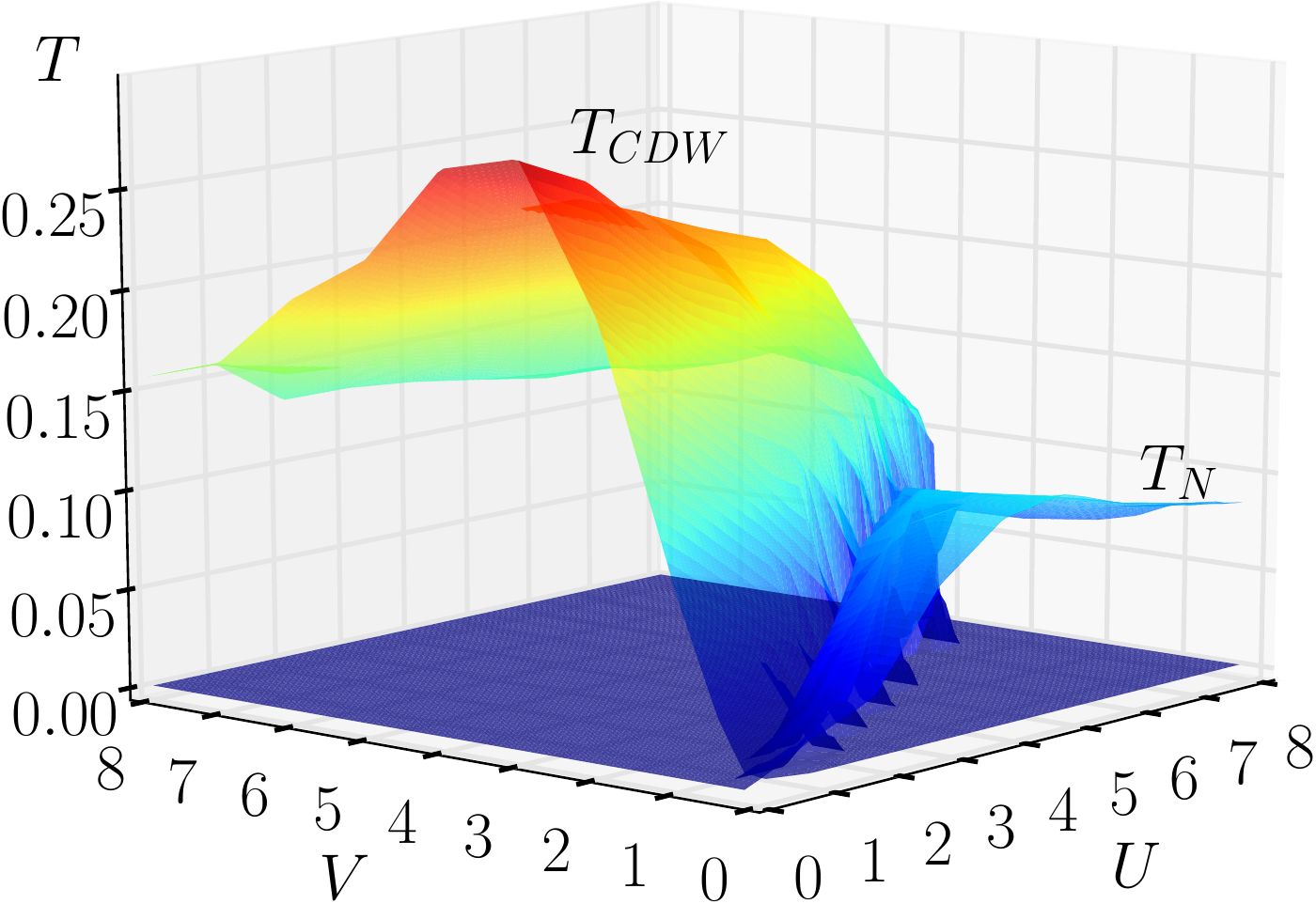}\\
\caption{(Color online) Transition temperature for the charge density wave and antiferromagnetic phase for different values of $U$ and $V$. }
\label{fig:fig9}
 \end{figure} 
 
  \section{Conclusions}

We presented above a numerical study of the static Holstein-Hubbard model by employing Hubbard-Stratanovich auxiliary fields for the charge and spin sectors. It captures many of the features obtained in previous DMF studies. More importantly,  it sheds light on new physics at finite temperatures and intermediate couplings due to the inclusion of spatial dependence (unlike DMFT) and classical thermal fluctuations through configuration sampling. The method works very well at all strengths of coupling. Being a real-space method it allows one to visualize the phases at various temperatures and how different orders develop and transform into others. It is numerically more efficient and large system sizes can be accessed.  Various physical properties such as single particle spectral functions, phonon distributions, and transport can be readily evaluated.\\
~\\
The salient results include the appearance of a nonmagnetic metallic phase at low values of coupling parameters, which was also seen in previous DMFT study, in addition to the ordered phases, including antiferromagnetic and charge-modulated ones. However, the finite-temperature phase diagram shows rich features and includes a pseudogap  phase at intermediate coupling. This arises due to the persistence of local order in charge and spin degrees. Inclusion of spatial correlations is essential to capture this region. However, a couple of remarks are in order. First, the ground-state transition from AF to CO is expected to be a first-order one. However, we find that this is a very weak transition and we are not able to resolve it accurately within the numerical error bars. The weak nature of this transition was also noted in previous DMFT studies. The low-temperature insulating states at small couplings could be a result of the fact that we used a 
two-dimensional square lattice. This necessarily gives a nesting instability at half filling and results in Slater or Peirels transition at low temperatures. Inclusion of quantum fluctuations or use of different  lattice geometries may obscure these phases. Indeed, DMFT study shows that the metallic phase exists at low strengths even when one of them is zero. Third, one could wonder whether these transitions are numerical artifacts since we have used a two-dimensional system. Since thermal fluctuations destroy any order at nonzero temperatures, we expect $T_N =$ 0 and $T_{CDW} =$ 0. However, it is expected that there would be a coherence temperature roughly mimicking the above transition temperatures even in two dimensions below which the correlation lengths increase rapidly. In other words,  the system enters the renormalized classical regime\cite{borejsza1}. If so, even a weak coupling  to a third dimension will stabilize the ordered phases. There could be some qualitative changes such as disappearance of insulating phases at weak couplings since nesting is no longer possible, but we expect gross features to remain the same.\\
~\\
The method presented neglects time dependence in auxiliary fields and phonons. Comparison of our results with previous DMFT studies suggests that quantum dynamical effects may not be highly relevant for these phases especially since the system orders at low temperatures. However, this is indeed a handicap and does not allow us to explore other instabilities such as superconductivity. The present method may be thought of in the same spirit as spin wave theory applied to spin systems, whereby  one starts with a classical ground state configuration for the spins and  builds up quantum corrections perturbatively in some small parameter. The present method is a step towards implementing such a procedure for 
many constituents interacting among themselves. However, it goes beyond the analogy of spin wave theory in several aspects.  We do not need to assume a classical ground state;
the Monte Carlo procedure  selects it naturally. The latter also incorporates thermal fluctuations. The procedure handles the weak to strong coupling limits in a unified way by smoothly interpolating between them. A further approximation is made by treating the charge fields at the saddle-point level. 
The charge fields ($\phi$) turn out to be purely imaginary  and their contribution to the classical action does not have a lower bound.
This makes the classical Monte Carlo sampling  of these fields very unstable and necessitates the saddle-point approximation which we have resorted to. The results obtained this way, for the pure Hubbard and Holstein models, respectively (see Fig. (1)), suggest that the saddle-point approximation does not affect the results severely. Finally, since quantum fluctuations of the
auxiliary fields are neglected, the results become less reliable at low temperatures and lower dimensions, where quantum effects may either destroy or modify the classical ground state
as happens in spin systems. A natural way of capturing corrections would be to allow small amplitude fluctuation of the classical variables around their equilibrium value at a Gaussian level and look at the stability of phases. That necessitates a new line of study and we postpone it for the future  \cite{subrat}.\\
~\\
The current work is not aimed at probing the physics of various families of experimentally accessible systems in which phonon dynamics is crucial along with electron
correlation effects. Instead, we attempted to study a model system using some physically appealing simplifications.  However, we believe that the results would  provide us some
insight into the behavior of systems in which relevant phonon frequencies are very large. Let us look at some relevant experimental systems.  For the cuprate family \cite{damascelli, lanzara},
$t \sim$ 0.25 eV, $U \sim$ 3 eV, $\omega \sim$ 75 meV, and the dimensionless electron-phonon coupling constant $V \sim$ 1. Thus the scaled Hubbard interaction is roughly three times the electron-phonon coupling and at half filling they would be Mott insulators with no charge-ordering tendencies. However, the phonon dynamics is relevant since
the relevant frequencies are not very large.  For fullerides, bandwidth is roughly 0.6 eV, $U \sim$ 1 eV,  $\omega \sim$ 90 meV, and $V$ lies in the range of 0.5-1 \cite{capone,
gunnersen}. Here again, Coulomb interactions dominate, though phonon dynamics is crucial. The family of manganites presents a more complex and rich scenario \cite{rama,venkat}.
The parameter values relevant for these systems turn out to be $t \sim$ 0.2-0.4 eV, $U \sim$ 3-4 eV, $\omega \sim$ 0.05 eV (the energy of the Jahn-Teller phonons), while
the Jahn-Teller polaronic energy $E_{JT} \sim$ 0.5-1 eV. While the scaled electron-phonon coupling $V$ is large (in the range of 1-4), the adiabaticity parameter
$\gamma = \hbar \omega/E_F \sim$ 0.2-0.3, where $E_F$ is the Fermi energy, and hence small. Naturally these are candidates more relevant for the present study, albeit at half filling.
However, the physics gets more complicated due to a variety of reasons: multiorbital and multi-(JT)-phonon effects and the Hund's rule coupling, not to mention the cooperative nature of the JT modes and associated charge-ordering tendencies. A generalization of our method should be appropriate to studying these systems.\\
~\\
The method can be expanded to study many problems of current interest. We mention a couple of them. In the present study we have limited ourselves to  a single-band Hubbard model coupled to a single-phonon mode. Many interesting realistic systems, such as manganites\cite{rama,Dagottobook} and iridates\cite{iridates1}, involve multiorbitals and multiphonon modes. However, the present method can be generalized naturally to include them. The computational complexity increases marginally, but the problem is tractable within the approximations used. The method could also be extended to study interfaces\cite{millis2} and/or heterostructures\cite{Anand} of correlated, electron-phonon problems which are difficult to handle in conventional methods that are being currently used.\\
~\\
A central feature of the method lies in capturing spatial correlations. This is essential for capturing features arising due to local order. More importantly, methods such as DMFT that neglect spatial dependence are not suited to study geometries\cite{iridates1,organic,kagome} where spatial features are important. A relevant case is the Holstein-Hubbard physics in frustrated geometries.
We are currently pursuing this problem which shows rich physics including transition from charge-ordered phases to charge stripes, nontrivial spin orders, etc. These results will be presented elsewhere\cite{pradhan}.\\
~\\

\section{Acknowledgements}
We acknowledge the use of  high performance computing clusters at HRI. We thank the anonymous referees for their constructive comments and suggestions.

\end{document}